\documentclass[12pt,a4paper,english,onecolumn,draftcls]{IEEEtran}
\usepackage[T1]{fontenc}
\usepackage[latin9]{luainputenc}
\usepackage{xcolor}
\usepackage{verbatim}
\usepackage{amsmath}
\usepackage{amsthm}
\usepackage{amssymb}
\usepackage{graphicx}
\usepackage{setspace}
\makeatletter

\@ifundefined{pageheight}{\let\pageheight\pdfpageheight}{}
\@ifundefined{pagewidth}{\let\pagewidth\pdfpagewidth}{}
\pageheight\paperheight
\pagewidth\paperwidth

\providecommand{\tabularnewline}{\\}

\theoremstyle{plain}
\newtheorem{thm}{\protect\theoremname}
\theoremstyle{plain}
\newtheorem{lem}[thm]{\protect\lemmaname}

\usepackage{subfigure}
\usepackage{epstopdf}
\usepackage{cite}
\usepackage{citesort}
\usepackage{balance}

\makeatother

\usepackage{babel}
\providecommand{\lemmaname}{Lemma}
\providecommand{\theoremname}{Theorem}

\begin{document}

\title{
}

\title{Optimum Interference Management in Underlay Inband D2D-Enhanced Cellular
Networks }

\author{\noindent 
{\normalsize{}Junnan Yang$^{\dagger}$, Ming Ding$^{\ddagger}$, Guoqiang
Mao$^{\nparallel}$}\emph{\normalsize{} }\\
\textit{\footnotesize{}$^{\dagger}$$^{\nparallel}$School of Computing
and Communication, University of Technology Sydney, Australia,}\\
\textit{\footnotesize{}$^{\ddagger}$Data61, CSIRO, Australia, }\\
}
\maketitle
\begin{abstract}
For device-to-device (D2D) communications underlaying a cellular network
with uplink resource sharing, both cellular and D2D links cause significant
co-channel interference. In this paper, we address the critical issue
of interference management in the network considering a practical
path loss model incorporating both line-of-sight (LoS) and non-line-of-sight
(NLoS) transmissions. To reduce the severe interference caused by
active D2D links, we consider a mode selection scheme based on the
maximum received signal strength (MRSS) for each user equipment (UE)
to control the D2D-to-cellular interference. Specifically, a UE will
operate in a cellular mode, only if its received signal strength from
the strongest base station (BS) is larger than a threshold $\beta$;
otherwise, the UE will operate in a D2D mode. Furthermore, we analyze
the performance in terms of the coverage probability and the area
spectral efficiency (ASE) for both the cellular network and the D2D
one. Analytical results are obtained and the accuracy of the proposed
analytical framework is validated through Monte Carol simulations.
Through our theoretical and numerical analyses, we quantify the performance
gains brought by D2D communications in cellular networks and we find
an optimum mode selection threshold $\beta$ to maximize the total
ASE in the network.

\end{abstract}


\section{Introduction\label{sec:Introduction}}

In the last decade, there has been a sharp increase in the demand
for data traffic~\cite{index2016global}. To address such massive
consumer demand for data communications, especially from the powerful
user equipment (UEs) such as smartphones and tablets, several noteworthy
technologies have been proposed~\cite{7126919}, such as small cell
networks (SCNs), cognitive radio, device-to-device (D2D) communications,
etc. In particular, D2D communications allow direct data transfer
between a pair of neighboring mobile UEs. Due to the short communication
distance between such pairs of D2D UEs, D2D communications hold great
promise in improving network performance such as coverage, spectral
efficiency, energy efficiency and so on~\cite{6970763}.

In the standardization of the 5-th generation (5G) networks, the orthogonal
frequency division multiple access (OFDMA) based D2D communications
adopt two types of spectrum sharing methods, (i) inband (e.g., using
cellular spectrum) or (ii) outband (e.g., unlicensed spectrum). In
particular, in the inband D2D communications, D2D users can setup
their communications in an underlay or overlay manner. More specifically,
in an underlay setting, D2D users access the same spectrum of cellular
users (CUs) whereas in overlay, D2D users access a dedicated portion
of cellular spectrum~\cite{3gpp}. Recently, D2D underlaying cellular
networks have been standardized by the 3rd Generation Partnership
Project (3GPP)\cite{TR36.814}. For the underlay inband D2D communications,
the most critical issue is to reduce the interference as cellular
links and D2D links share the same radio resources.

Although the reuse of the cellular spectrum via D2D can improve the
area spectral efficiency of the network, such D2D operations also
pose great challenges. The major challenge in the D2D-enabled cellular
network is the existence of inter-tier and intra-tier interference
due to the aggressive frequency reuse, where cellular UEs and D2D
UEs share the same spectrum. It is essential to design an effective
interference management scheme to control the interference generated
by the D2D links to the cellular links, and vice versa. Consequently,
there has been a surge of academic studies in this area. Transmission
power control~\cite{6909030,6928445,7933260,lee2015power}, distance
based mode selection~\cite{7147772} and a guard zone interference
control scheme~\cite{6047553,7147834,7676388} have been proposed
to solve this problem. In this paper, we present a novel mode selection
scheme based on the maximum received signal strength for each user
equipment (UE) to control the interference.\textcolor{brown}{{} }In
more detail, a UE will operate in a cellular mode if its received
signal strength from the strongest base station (BS) is larger than
a threshold $\beta$ ; otherwise, it will operate in a D2D mode. This
will mitigate large interference from the D2D links to the cellular
links.\textcolor{brown}{{} }To analyze the proposed interference control
scheme, we develop a theoretical framework that takes power control,
practical path loss and lognormal fading into account. Based on our
analytical results, we find a tradeoff between the maximization of
the ASE performance and the fairness of the D2D links, and the optimum
setting of the threshold $\beta$ that maximizes the ASE.

Moreover, the path loss models of D2D links and cellular links in
a D2D-enabled cellular network are different due to the difference
in the heights and the locations of transmitters~\cite{our_work_TWC2016}.
It is well known that LoS transmission may occur when the distance
between a transmitter and a receiver is small, and NLOS transmission
is common in office environments and in central business districts.
Furthermore, when the distance between a transmitter and a receiver
decreases, the probability that a LoS path exists between them increases,
thereby causing a transition from NLoS transmission to LoS transmission
with a higher probability. Due to the proximity between D2D users,
the physical channels which constitute D2D communications are expected
to be complex in nature, experiencing both LoS and NLoS conditions
across these pairs, which are distinctly different from conventional
cellular environments~\cite{7890358}. Generally speaking, D2D links
are more likely to operate in LoS conditions while the cellular links
are more likely to operate in NLoS conditions. To the best of our
knowledge, there have been no studies that investigate network performance
of D2D enhanced cellular networks, which adopts different path loss
models for the cellular links and the D2D links. Our analysis shows
non-trivial difference on the network performance when considering
different path loss models for the cellular links and the D2D links
respectively, which captures the different environment conditions
that cellular links and D2D links operate in.

Compared with the existing work, the main contributions of this paper
are:
\begin{itemize}
\item We proposed a tractable interference management scheme for each user
equipment (UE) to control the co-channel interference. Specifically,
a UE will operate in a cellular mode if its received signal strength
from the strongest base station (BS) is larger than a threshold $\beta$
; otherwise, it will operate in a D2D mode.
\item We present a general analytical framework using stochastic geometry
and intensity matching approach~\cite{7482733}. Then, we derive
the results of coverage probability and ASE for both the cellular
mode and the D2D mode UEs. Our framework considers interference management,
LoS/NLoS transmission and shadow fading. The accuracy of our analytical
results is validated by Monte Carlo simulations.
\item Different from the existing work that does not differentiate the path
loss models between cellular links and D2D links, our analysis adopts
two different path loss models for cellular links and D2D links, respectively.
Our results demonstrate that the D2D links can provide a considerable
ASE gain when the threshold parameter is appropriately chosen. More
specifically, our analysis shows the interference from D2D tier can
be controlled by using our mode selection scheme, and there is an
optimal $\beta$ to achieve the maximum ASE while the performance
of cellular tier is guaranteed.
\end{itemize}
The rest of this paper is structured as follows. Section~\ref{sec:Related-Work}
provides a brief review of related work. Section~\ref{sec:System-Model}
describes the system model. Section~\ref{sec:General-Results} presents
our theoretical analysis on the coverage probability and the ASE with
applications in a 3GPP special case. The numerical and simulations
results are discussed in Section~\ref{sec:Simulation-and-Discussion}.
Our conclusions are drawn in Section~\ref{sec:Conclusion}.

\section{Related Work\label{sec:Related-Work}}

D2D communications underlaying cellular networks are ongoing standardization
topics in LTE-A~\cite{TR36.828}. Meanwhile, stochastic geometry
which is accurate in modeling irregular deployment of base stations
(BSs) and mobile user equipment (UEs) has been widely used to analyze
network performance~\cite{6042301,6516885,peng2014device}. Andrews,
et.al conducted network performance analyses for the downlink (DL)~\cite{6042301}
and the uplink (UL)~\cite{6516885} of SCNs, in which UEs and/or
base stations (BSs) were assumed to be randomly deployed according
to a homogeneous Possion point process (HPPP). In~\cite{peng2014device},
Peng developed an analytical framework for the D2D communications
underlaied cellular network in the DL, where a Rician fading channel
model was adopted to model the small-scale fast fading for the D2D
communication links. Although some studies assumed that D2D links
operate on the DL spectrum, and hence the interference from BSs to
D2D receivers is severe. In practice, allowing D2D links to access
the UL spectrum might be a more realistic assumption, as 3GPP has
standardized D2D communications~\cite{36.877}\textcolor{brown}{.}

On the other hand, as one of the fundamental performance metrics of
the communication system, D2D transmission capacity has been analyzed
in the literature~\cite{6047553,6909030,6928445,7147772,7147834,7676388}.
In~\cite{6047553}, the author proposed an interference-limited area
control scheme to mitigate the interference from cellular to D2D considering
a single slope path loss model. In~\cite{lee2015power}, Lee proposed
a power control algorithm to control the co-channel interference in
which global channel state information are required at BSs. In~\cite{7147772},
Liu provided a unified framework to analyze the downlink outage probability
in a multi-channel environment with Rayleigh fading, where D2D UEs
were selected based on the average received signal strength from the
nearest BS, which is equivalent to a distance-based selection. The
authors of~\cite{7147834} and~\cite{7676388} proposed novel approaches
to model the interference in uplink or downlink underlaid/overlaid
with Rayleigh fading and single path loss model.

Meanwhile, limited studies have been conducted to consider D2D networks
with general fading channels, for example in~\cite{7890358} and~\cite{peng2014device},
the authors considered generalized fading conditions and analyzed
the network performance, while they did not differentiate the path
loss models between the D2D links and cellular links.

Although the existing works have provided precious insights into resource
allocation and capacity enhancement for D2D communications, there
are several remaining problems:
\begin{itemize}
\item The mode selection schemes in the literatures were not very practical,
they were mostly based on the UE-to-BS distance while a more practical
one which based on the maximum received signal strength should be
considered.
\item In some studies, only a single BS with one cellular UE and one D2D
pair were considered, which did not take into account the influence
from other cells. Moreover, in most studies, the authors considered
D2D receiver UEs as an additional tier of nodes, independent of the
cellular UEs and the D2D transmitter UEs. Such tier of D2D receiver
UEs without cellular capabilities appears from nowhere and is hard
to justify in practice.
\item The path loss model is not practical, e.g., the impact of LoS/NLoS
conditions have not been well studied in the context of D2D and usually
the same path loss model was used for both the cellular and the D2D
tiers. In addition, shadow fading was widely ignored in the existing
analyses, which did not reflect realistic networks.
\end{itemize}

To sum up, in this paper, we propose a more generalized framework
which takes into account a novel interference management scheme based
on the maximum received signal strength, probabilistic NLoS and LoS
transmissions and lognormal shadow fading, and shed new insight on
the interference management of coexistent D2D and cellular transmissions.

\section{System Model\label{sec:System-Model}}

In this section, we first explain the scenario of the D2D communication
coexisting with cellular network. Then, we present the path loss model
and the mode selection scheme.

\subsection{Scenario Description\label{subsec:Scenario-Description}}

We consider a D2D underlaid UL cellular network, where BSs and UEs,
including cellular UL UEs and D2D UEs, are assumed to be distributed
on an infinite two-dimensional plane $\mathbf{\mathit{\mathtt{\mathbb{R}}}^{2}}$.
We assume that the cellular BSs are spatially distributed according
to a homogeneous PPP of intensity $\lambda_{b}$ , i.e., $\varPhi_{b}=\{X_{i}\}$,
where $X_{i}$ denotes the spatial locations of the $i$th BS. Moreover,
the UEs are also distributed in the network region according to another
independent homogeneous PPP $\varPhi_{u}$ of intensity $\lambda_{u}$.

\subsection{Path Loss Model\label{subsec:Path-Loss-Model}}

We incorporate both NLoS and LoS transmissions into the path loss
model. Following~\cite{our_GC_paper_2015_HPPP,our_work_TWC2016},
we adopt a very general path loss model, in which the path loss $\zeta\left(r\right)$,
as a function of the distance $r$, is segmented into $N$ pieces
written as%
\begin{equation}
\zeta\left(r\right)=\begin{cases}
\zeta_{1}\left(r\right), & \textrm{when }0\leq r\leq d_{1}\\
\zeta_{2}\left(r\right), & \textrm{when }d_{1}<r\leq d_{2}\\
\vdots & \vdots\\
\zeta_{N}\left(r\right), & \textrm{when }r>d_{N-1}
\end{cases},\label{eq:prop_PL_model}
\end{equation}
where each piece $\zeta_{n}\left(r\right),n\in\left\{ 1,2,\ldots,N\right\} $
is modeled as
\begin{equation}
\zeta_{n}\left(r\right)\hspace{-0.1cm}=\hspace{-0.1cm}\begin{cases}
\hspace{-0.2cm}\begin{array}{l}
\zeta_{n}^{\textrm{L}}\left(r\right)=A_{n}^{{\rm {L}}}r^{-\alpha_{n}^{{\rm {L}}}},\\
\zeta_{n}^{\textrm{NL}}\left(r\right)=A_{n}^{{\rm {NL}}}r^{-\alpha_{n}^{{\rm {NL}}}},
\end{array} & \hspace{-0.2cm}\hspace{-0.3cm}\begin{array}{l}
\textrm{LoS Probability:}~\textrm{Pr}_{n}^{\textrm{L}}\left(r\right)\\
\textrm{NLoS Probability:}~1-\textrm{Pr}_{n}^{\textrm{L}}\left(r\right)
\end{array}\hspace{-0.1cm},\end{cases}\label{eq:PL_BS2UE}
\end{equation}
where
\begin{itemize}
\item $\zeta_{n}^{\textrm{L}}\left(r\right)$ and $\zeta_{n}^{\textrm{NL}}\left(r\right),n\in\left\{ 1,2,\ldots,N\right\} $
are the $n$-th piece path loss functions for the LoS transmission
and the NLoS transmission, respectively,
\item $A_{n}^{{\rm {L}}}$ and $A_{n}^{{\rm {NL}}}$ are the path losses
at a reference distance $r=1$ for the LoS and the NLoS cases, respectively,
\item $\alpha_{n}^{{\rm {L}}}$ and $\alpha_{n}^{{\rm {NL}}}$ are the path
loss exponents for the LoS and the NLoS cases, respectively.
\end{itemize}
\noindent In practice, $A_{n}^{{\rm {L}}}$, $A_{n}^{{\rm {NL}}}$,
$\alpha_{n}^{{\rm {L}}}$ and $\alpha_{n}^{{\rm {NL}}}$ are constants
obtainable from field tests and continuity constraints~\cite{SCM_pathloss_model}.

As a special case, we consider a path loss function adopted in the
3GPP~\cite{TR36.828}, and we adopt two different path loss models
for cellular links and D2D links as

\begin{equation}
\zeta_{B}\left(r\right)\hspace{-0.1cm}=\hspace{-0.1cm}\begin{cases}
\hspace{-0.2cm}\begin{array}{l}
A_{B}^{{\rm {L}}}r^{-\alpha_{B}^{{\rm {L}}}},\\
A_{B}^{{\rm {NL}}}r^{-\alpha_{B}^{{\rm {NL}}}},
\end{array} & \hspace{-0.2cm}\hspace{-0.3cm}\begin{array}{l}
\textrm{LoS Probability:}~\textrm{Pr}_{B}^{\textrm{L}}\left(r\right)\\
\textrm{NLoS Probability:}~1-\textrm{Pr}_{B}^{\textrm{L}}\left(r\right)
\end{array}\hspace{-0.1cm},\end{cases}\label{eq:PL_BS2UEspecial case-1}
\end{equation}
and

\begin{equation}
\zeta_{D}\left(r\right)\hspace{-0.1cm}=\hspace{-0.1cm}\begin{cases}
\hspace{-0.2cm}\begin{array}{l}
A_{D}^{{\rm {L}}}r^{-\alpha_{D}^{{\rm {L}}}},\\
A_{D}^{{\rm {NL}}}r^{-\alpha_{D}^{{\rm {NL}}}},
\end{array} & \hspace{-0.2cm}\hspace{-0.3cm}\begin{array}{l}
\textrm{LoS Probability:}~\textrm{Pr}_{D}^{\textrm{L}}\left(r\right)\\
\textrm{NLoS Probability:}~1-\textrm{Pr}_{D}^{\textrm{L}}\left(r\right)
\end{array}\hspace{-0.1cm},\end{cases}\label{eq:PL_BS2UEspecial case-2}
\end{equation}
together with a linear LoS probability function as follows~\cite{TR36.828},
\begin{equation}
\textrm{Pr}_{B}^{\textrm{L}}\left(r\right)=\begin{cases}
1-\frac{r}{d_{B}} & 0<r\leq d_{B}\\
0 & r>d_{B}
\end{cases},\label{eq:LoS probability function-2}
\end{equation}
and
\begin{equation}
\textrm{Pr}_{D}^{\textrm{L}}\left(r\right)=\begin{cases}
1-\frac{r}{d_{D}} & 0<r\leq d_{D}\\
0 & r>d_{D}
\end{cases},\label{eq:LoS probability function-1}
\end{equation}
where $d_{B}$ and $d_{D}$ is the cut-off distance of the LoS link
for UE-to-BS links and UE-to-UE links. The adopted linear LoS probability
function is very useful because it can include other LoS probability
functions as its special cases~\cite{our_work_TWC2016}.

\subsection{User Mode Selection Scheme\label{subsec:User-Mode-Selection}}

There are two modes for UEs in the considered D2D-enabled UL cellular
network, i.e., cellular mode and D2D mode. Each UE is assigned with
an operation mode according to the comparison of the maximum received
DL power from its serving BS with a threshold. In more detail, the
considered user model selection criterion is formulated as
\begin{equation}
Mode=\begin{cases}
\textrm{Cellular}, & \textrm{if }P^{\ast}=\underset{b}{\max}\left\{ P_{b}^{\textrm{rx}}\right\} >\beta\\
\textrm{D2D}, & \textrm{otherwise}
\end{cases},\label{eq:modeselction}
\end{equation}
where the string variable $Mode$ takes the value of 'Cellular' or
'D2D' to denote the cellular mode and the D2D mode, respectively.
In particular, for a tagged UE, if $P^{\ast}$ is large than a specific
threshold $\beta>0$. This UE is not appropriate to work in the D2D
mode due to its potentially large interference to cellular UEs. Hence,
it should operate in the cellular mode and directly connect with the
strongest BS; otherwise, it should operate in the D2D mode. The UEs
Which are associated with cellular BSs are referred to as cellular
UEs (CUs). The distance from a CU to its associated BS is denoted
by $R_{B}$. From~\cite{6928445}, we assume CUs are distributed
following a non-homogenous PPP $\varPhi_{c}$. For a D2D UE, we adopt
the same assumption in~\cite{7147772} that it randomly decides to
be a D2D transmitter or a D2D receiver with equal probability at the
beginning of each time slot, and a D2D receiver UE selects the strongest
D2D transmitter UE for signal reception.

The received power for a typical UE from a BS $b$ can be written
as
\begin{equation}
P_{b}^{\textrm{rx}}=\begin{cases}
A_{BL}P_{B}\mathrm{\mathcal{H}_{B}}\left(b\right)R_{B}^{-\alpha_{BL}} & \mathtt{\text{LoS}}\\
A_{BN}P_{B}\mathrm{\mathcal{H}_{B}}\left(b\right)R_{B}^{-\alpha_{BN}} & \textrm{otherwise}
\end{cases},\label{eq:maximumreceivedpower}
\end{equation}
where $A_{BL}=10^{\frac{1}{10}A_{BL}^{\textrm{dB}}}$ and $A_{BN}=10^{\frac{1}{10}A_{BN}^{\textrm{dB}}}$
denote a constant determined by the transmission frequency for BS-to-UE
links in LoS and NLoS conditions, respectively. $P_{B}$ is the transmission
power of a BS, $\mathrm{\mathcal{H}_{B}}\left(b\right)$ is the lognormal
shadowing from a BS $b$ to the typical UE. $\alpha_{BL}$ and $\alpha_{BN}$
denote the path loss exponents for BS-to-UE links with LoS and NLoS,
respectively. Base on the above system model, we can obtain the intensity
of CU as $\lambda_{c}=q\lambda_{u}$, where $q$ denotes the probability
of $P^{\ast}>\beta$ and will be derived in closed-form expressions
in Section~\ref{sec:General-Results}. It is apparent that the D2D
UEs are distributed following another non-homogenous PPP $\varPhi_{d}$,
the intensity of which is $\lambda_{d}=\left(1-q\right)\lambda_{u}$.
Considering that a required content file might not exist in a D2D
transmitter, in reality, we assume that $\rho\%$ D2D transmitters
possess the required content files and deliver them to D2D receivers.
In other words, $\rho\%$ of the D2D links will eventually work in
one time slot.

We assume an underlaid D2D model. That is, each D2D transmitter reuses
the frequency with cellular UEs, which incurs inter-tier interference
from D2D to cellular. However, there is no intra-cell interference
between cellular UEs since we assume an orthogonal multiple access
technique in a BS. It follows that there is only one uplink transmitter
in each cellular BS. Here, we consider a fully loaded network with
$\lambda_{u}\gg\lambda_{b}$, so that on each time-frequency resource
block, each BS has at least one active UE to serve in its coverage
area. Note that the case of $\lambda_{u}<\lambda_{b}$ is not trivial,
which even changes the capacity scaling law~\cite{Ding2017capScaling}.
In this paper, we focus on the former case, and leave the study of
$\lambda_{u}<\lambda_{b}$ as our future work. Generally speaking,
the active CUs can be treated as a thinning PPP $\varPhi_{c}$ with
the same intensity $\lambda_{b}$ as the cellular BSs.

Moreover, we assume a channel inversion strategy for the power control
for cellular UEs, i.e.,
\begin{equation}
P_{c_{i}}=\begin{cases}
P_{0}\mathcal{\mathrm{\left(\frac{R_{i}^{\alpha_{BL}}}{\mathcal{H_{\mathrm{c_{i}}}}A_{BL}}\right)^{\varepsilon}}} & \mathtt{\text{LoS}}\\
P_{0}\mathcal{\mathrm{\left(\frac{R_{i}^{\alpha_{BN}}}{\mathcal{H_{\mathrm{c_{i}}}}A_{BN}}\right)^{\varepsilon}}} & \text{otherwise}
\end{cases},\label{eq:cupowercontrol}
\end{equation}
where $P_{c_{i}}$ is the transmission power of the $i$-th UE in
cellular link, $R_{i}$ is the distance of the $i$-th link from a
CU to the target BS, $\mathcal{H_{\mathrm{c_{i}}}}$ is the lognormal
shadowing between target BS and the cellular UE, $\epsilon\in(0,1]$
is the fractional path loss compensation, $P_{0}$ is the receiver
sensitivity. For downlink BS and D2D transmitters, they use constant
transmit powers $P_{B}$ and $P_{d}$, respectively. Besides, we denote
the additive white Gaussian noise (AWGN) power by $\sigma^{2}$.

\subsection{Performance Metrics\label{subsec:The-Performance-Metrics}}

According to~\cite{6042301}, the coverage probability is defined
as
\begin{equation}
P_{Mode}\left(\gamma,\lambda_{u},\alpha_{B,D}\right)=\Pr\left[\textrm{SINR}>\gamma\right],\label{eq:definesinr}
\end{equation}
where $\gamma$ is the SINR threshold, the subscript string variable
$Mode$ takes the value of 'Cellular' or 'D2D', and the interference
in this paper consist of the interference from both cellular UEs and
D2D transmitters.

Furthermore, the area spectral efficiency (ASE) in $\textrm{bps/Hz/k\ensuremath{m^{2}}}$
can be formulated as
\begin{align}
 & A_{Mode}^{\textrm{ASE}}\left(\lambda_{Mode},\gamma_{0}\right)\label{eq:ase}\\
 & =\lambda_{Mode}\int_{\gamma_{0}}^{\infty}\log_{2}\left(1+x\right)f_{X}\left(\lambda_{Mode},\gamma_{0}\right)dx,\nonumber
\end{align}
where $\gamma_{0}$ is the minimum working SINR for the considered
network, and $f_{X}\left(\lambda_{Mode},\gamma_{0}\right)$ is the
PDF of the SINR observed at the typical receiver for a particular
value of $\lambda_{Mode}$.

For the whole network consisting of both cellular UEs and D2D UEs,
the sum ASE can be written as
\begin{equation}
A^{\textrm{ASE}}=A_{\textrm{Cellular}}^{\textrm{ASE}}+A_{\textrm{D2D}}^{\textrm{ASE}}.\label{eq:totalase}
\end{equation}

\section{Main Results\label{sec:General-Results}}

In this section, the performance of UEs are characterized in terms
of their coverage probability and ASE both for cellular tier and D2D
tier. The probability that the UE operates in the cellular mode is
derived in Section \ref{subsec:Coverage-Probability}, the coverage
probability of cellular UE and D2D UE are derived in Section \ref{subsec:Cellular-mode}
and Section \ref{subsec:Coverage-Probability-of}, respectively.

\subsection{Probability operating in the cellular mode\label{subsec:Probability-Operating-In}}

Due to consideration of lognormal shadowing in this mode we use the
intensity measure method in\cite{7482733} to first obtain an equivalent
network for further analysis. In particular, we transform the original
PPP with lognormal shadowing to a equivalent PPP which has the matched
intensity measure and intensity. More specifically, define $\overline{R}_{i}^{BL}=\mathrm{\mathcal{H}_{B}^{-1/\alpha_{BL}}}R_{i}^{BL}$
and $\overline{R}_{i}^{BN}=\mathrm{\mathcal{H}_{B}^{-1/\alpha_{BN}}}R_{i}^{BN}$,
where $R_{i}^{BL}$ and $R_{i}^{BN}$ are the distance separating
a typical user from its tagged strongest base station with LoS and
NLoS. $\overline{R}_{i}^{BL}$ and $\overline{R}_{i}^{BN}$ are the
equivalent distance separating a typical user from its tagged nearest
base station in the new PPP.

The network consists of two non-homogeneous PPPs with intensities
$\lambda p^{NL}(R_{i})$ and $\lambda p^{L}(R_{i})$, which representing
the sets of NLoS and LoS links respectively. Each UE is associated
with the strongest transmitter. Moreover, intensities $\lambda{}^{NL}(\cdot)$
and $\lambda{}^{L}(\cdot)$ are given by
\begin{equation}
\lambda{}^{NL}(t)=\frac{d}{dt}\varLambda^{NL}\left(\left[0,t\right]\right)
\end{equation}
and
\begin{equation}
\lambda{}^{L}(t)=\frac{d}{dt}\varLambda^{L}\left(\left[0,t\right]\right)
\end{equation}
respectively, where
\begin{equation}
\varLambda^{NL}\left(\left[0,t\right]\right)=\mathbb{E}_{\mathcal{H}}\left[2\pi\lambda\int_{0}^{t\left(\mathcal{H}\right)^{1/\alpha^{NL}}}p^{NL}(r)rdr\right]\label{eq:intensity nlos}
\end{equation}
and
\begin{equation}
\varLambda^{L}\left(\left[0,t\right]\right)=\mathbb{E}_{\mathcal{H}}\left[2\pi\lambda\int_{0}^{t\left(\mathcal{H}\right)^{1/\alpha^{L}}}p^{L}(r)rdr\right].\label{eq:intensity los}
\end{equation}

Similar definition are adopted to D2D tier as well. The transformed
network has the exactly same performance for the typical receiver
(BS or D2D RU) on the coverage probability with the original network.

In this subsection, we present our results on the probability that
the UE operates in the cellular mode and the equivalence distance
distributions in the cellular mode and D2D mode, respectively. In
the following, we present our first result in Lemma~\ref{lem:When-operating-under},
which will be used in the later analysis of the coverage probability.
\begin{lem}
\label{lem:When-operating-under}The probability that a typical UE
connects to the strongest BS and operates in the cellular mode $q$
is given by
\begin{align}
q & =1-\exp\left[-\mathbb{E}_{\mathcal{H}}\left[2\pi\lambda_{B}\int_{0}^{\left(\frac{P_{b}\textrm{A}_{BL}\mathcal{H}}{\beta}\right)^{1/\alpha_{BL}}}p^{L}(r)rdr\right]\right.\nonumber \\
 & -\left.\mathbb{E}_{\mathcal{H}}\left[2\pi\lambda_{B}\int_{0}^{\left(\frac{P_{b}\textrm{A}_{BN}\mathcal{H}}{\beta}\right)^{1/\alpha_{BN}}}p^{NL}(r)rdr\right]\right],\label{eq:q}
\end{align}
 and the probability that the UE operates in the D2D mode is $\left(1-q\right)$.
\end{lem}
\begin{IEEEproof}
See Appendix A.%
\end{IEEEproof}
Note that Eq.(\ref{eq:q}) explicitly account for the effects of shadow
fading, pathloss, transmit power, spatial distribution of BSs and
mode selection threshold $\beta$ . From the result, one can see that
the HPPP $\phi_{u}$ can be divided into two PPPs: the PPP with intensity
$q\lambda_{u}$ and the PPP with intensity $(1-q)\lambda_{u}$, which
representing cellular UEs and D2D UEs, respectively. Same as in~\cite{6928445},
We assume these two PPPs are independent.

\begin{figure}
\begin{centering}
\includegraphics[width=8.8cm]{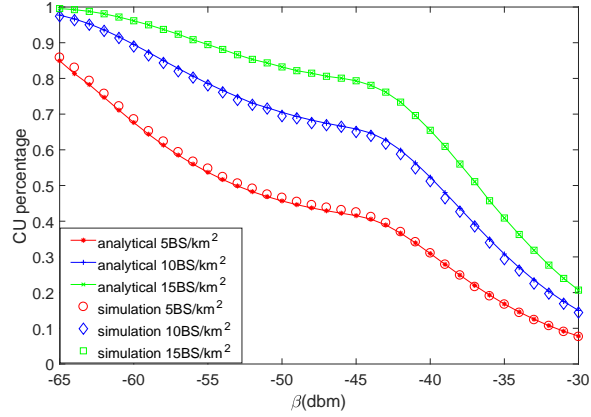}
\par\end{centering}
\caption{\label{fig1}The probability for a UE to operate in the cellular model
vary the RSS threshold $\beta$ , $\textrm{P}_{B}=46\textrm{dBm}$,
log-normal shadowing with zero means, $\sigma_{B}^{2}=8\text{dB}$
and $\sigma_{D}^{2}=7\text{dB}$}
\end{figure}

Fig.1 illustrates the probability for a UE to operate in the cellular
model based on Eq.(\ref{eq:q}). It can be seen that the simulation
results perfectly match the analytical results. From Fig.1, we can
find that over 50\% UEs can operate in the cellular mode when $\beta$
is smaller than -55 dBm as the BS intensity is 5$\text{BS/k\ensuremath{m^{2}}}$.
This value increases by approximately to -37 dBm and -35 dBm when
the BS intensity is 10$\mathtt{\text{BS/k\ensuremath{m^{2}}}}$ and
15$\text{BS/k\ensuremath{m^{2}}}$, respectively. It indicates that
the percentage of CUs will increase as the BS intensity grows.

\subsection{Coverage probability \label{subsec:Coverage-Probability}}

In this subsection, we investigate the coverage probability that a
receiver's signal-to-interference-plus-noise ratio (SINR) is above
a per-designated threshold $\gamma$:

\begin{equation}
P_{Mode}\left(T,\lambda_{u},\alpha_{B,D}\right)=\Pr\left[\textrm{SINR}>\gamma\right]\label{eq:definesinr-1}
\end{equation}

\noindent where $\gamma$ is the SINR threshold, the subscript string
variable $Mode$ takes the value of 'Cellular' or 'D2D', the SINR
is calculated as

\begin{equation}
\mathrm{SINR}=\frac{P_{Mode}\zeta_{Mode}\left(r\right)\mathcal{H_{\mathit{Mode}}}}{I_{cellular}+I_{d2d}+N_{0}},\label{eq:SINR defined}
\end{equation}
where $\mathcal{H_{\mathrm{Mode}}}$ is the lognormal shadowing between
transmitter and receiver in cellular mode or D2D mode. $P_{B}$, $P_{D}$
and $N_{0}$ are the transmission power of each cellular and D2D UE
transmitter and the additive white Gaussian noise (AWGN) power at
each receiver, respectively. $I_{cellular}$ and $I_{d2d}$ is the
cumulative interference given by $I_{cellular}=\sum_{i:\,c_{i}\in\Phi_{c}\setminus signal}P_{c,i}\beta_{i}\mathcal{H}_{i},$
and $I_{d2d}=\sum_{j:\,d_{i}\in\Phi_{d2d}\setminus signal}P_{D}\beta_{j}\mathcal{H}_{j},$
where $c_{i}$ and $d_{j}$ are the $i$-th interfering CU and $j$-th
interfering TU, $P_{c,i}$ is the transmit power $i$-th interfering
CU, $\beta_{i}$ ,$\beta_{j}$ and $\mathcal{H_{\mathrm{i}}}$, $\mathcal{H_{\mathrm{j}}}$
are the path loss associated with $c_{i}$ and $d_{j}$, and the lognormal
fading associated with $c_{i}$ and $d_{j}$, respectively.

\subsubsection{Coverage probability of cellular mode\label{subsec:Cellular-mode}}

Based on the path loss model in Eq.(\ref{eq:PL_BS2UEspecial case-1},\ref{eq:LoS probability function-2})
and the equivalence method in subsection~\ref{subsec:Probability-Operating-In},
we present our main result on $p_{c}^{\textrm{cov}}\left(\lambda,\gamma\right)$
in Theorem~\ref{thm:coverage of cellular mode}.
\begin{thm}
\noindent {\small{}\label{thm:coverage of cellular mode}}For the
typical BS which is located at the origin, considering the path loss
model in Eq.(\ref{eq:PL_BS2UEspecial case-1}) and the equivalence
method, the coverage probability $p_{c}^{{\rm {cov}}}\left(\lambda,\gamma\right)$
can be derived as
\begin{equation}
p_{c}^{{\rm {cov}}}\left(\lambda,\gamma\right)=T_{c}^{{\rm {L}}}+T_{c}^{{\rm {NL}}},\label{eq:Theorem_1_p_cov}
\end{equation}
where $T_{c}^{{\rm {L}}}=\int_{0}^{t_{LoS}}\left(\mathcal{\int_{\mathrm{-}\infty}^{\infty}\mathrm{\left[\frac{1-e^{-i\omega/\gamma}}{2\pi i\omega}\right]\mathcal{F}_{\frac{\textrm{1}}{SINR^{L}}}(\omega)}}d\omega\right)f_{\overline{R_{LCU}}}(r)dr$
and

\noindent $T_{c}^{{\rm {NL}}}=\int_{0}^{t_{NLoS}}\left(\mathcal{\int_{\mathrm{-}\infty}^{\infty}\mathrm{\left[\frac{1-e^{-i\omega/\gamma}}{2\pi i\omega}\right]\mathcal{F}_{\frac{\textrm{1}}{SINR^{NL}}}(\omega)}}d\omega\right)f_{\overline{R_{NLCU}}}(r)dr$
,

\noindent $t_{LoS}=\left(\frac{\beta}{\textrm{\ensuremath{P_{B}A^{L}}}}\right){}^{-1/\alpha_{BL}}$,$t_{NLoS}=\left(\frac{\beta}{\textrm{\ensuremath{P_{B}A^{NL}}}}\right){}^{-1/\alpha_{BN}}$,

\noindent $f_{\overline{R_{LCU}}}(r)$ and $f_{\overline{R_{NLCU}}}(r)$
, are represented by
\begin{equation}
f_{\overline{R_{LCU}}}^{{\rm {L}}}\left(r\right)=\frac{\exp\left(\hspace{-0.1cm}-\hspace{-0.1cm}\int_{0}^{\overline{r_{1}}}\left({\rm {Pr}}^{{\rm {NL}}}\left(u\right)\right)\lambda_{B}^{NL}(u)du\right)\exp\left(\hspace{-0.1cm}-\hspace{-0.1cm}\int_{0}^{r}{\rm {Pr}}^{{\rm {L}}}\left(u\right)\lambda_{B}^{L}(u)du\right){\rm {Pr}}^{{\rm {L}}}\left(r\right)\lambda_{B}^{L}(r)}{q},\label{eq:geom_dis_PDF_UAS1_LoS_thm}
\end{equation}
and
\begin{equation}
f_{\overline{R_{NLCU}}}^{{\rm {NL}}}\left(r\right)=\frac{\exp\left(\hspace{-0.1cm}-\hspace{-0.1cm}\int_{0}^{\overline{r_{2}}}{\rm {Pr}}^{{\rm {L}}}\left(u\right)\lambda(u)du\right)\exp\left(\hspace{-0.1cm}-\hspace{-0.1cm}\int_{0}^{r}\left({\rm {Pr}}^{{\rm {NL}}}\left(u\right)\right)\lambda_{B}^{NL}(u)du\right){\rm {Pr}}^{{\rm {NL}}}\left(r\right)\lambda_{B}^{NL}(r)}{q},\label{eq:geom_dis_PDF_UAS1_NLoS_thm}
\end{equation}
\end{thm}
\noindent where $\overline{r_{1}}$ and $\overline{r_{2}}$ are given
implicitly by the following equations as
\begin{equation}
\overline{r_{1}}=\underset{\overline{r_{1}}}{\arg}\left\{ \zeta^{{\rm {NL}}}\left(\overline{r_{1}}\right)=\zeta_{n}^{{\rm {L}}}\left(\overline{r}\right)\right\} ,\label{eq:def_r_1}
\end{equation}

\noindent and
\begin{equation}
\overline{r_{2}}=\underset{\overline{r_{2}}}{\arg}\left\{ \zeta^{{\rm {L}}}\left(\overline{r_{2}}\right)=\zeta_{n}^{{\rm {NL}}}\left(\overline{r}\right)\right\} .\label{eq:def_r_2}
\end{equation}

\noindent In addition, $\mathcal{F}_{\frac{\textrm{1}}{SINR^{L}}}(\omega)$
and $\mathcal{F}_{\frac{\textrm{1}}{SINR^{NL}}}(\omega)$ are respectively
computed by
\begin{align}
\mathcal{F}_{\frac{\textrm{1}}{SINR^{L}}}(\omega) & =\exp\left(-\int_{r}^{\infty}\left(1-\int_{0}^{t_{LoS}}\left[\exp\left(\mathrm{i\omega\frac{\mathrm{\left(z^{\alpha_{BL}}\right)^{\varepsilon}}v^{-\alpha_{BL}}}{A_{BL}^{2\epsilon}\left(r^{-\alpha^{BL}}\right)^{1-\varepsilon}}}\right)\right]f_{\overline{R_{LCU}}}(z)dz\right)\lambda_{B}^{L}(v)dv\right)\nonumber \\
\times & \exp\left(-\int_{r}^{\infty}\left(1-\int_{0}^{t_{LoS}}\left[\exp\left(\mathrm{i\omega\frac{\mathcal{\mathrm{\left(\frac{z^{\alpha_{BL}}}{A_{BL}}\right)^{\varepsilon}}}A_{BN}v^{-\alpha_{BN}}}{\left(A_{BL}r^{-\alpha^{BL}}\right)^{1-\varepsilon}}}\right)\right]f_{\overline{R_{LCU}}}(z)dz\right)\lambda_{B}^{NL}(v)dv\right)\nonumber \\
\times & \exp\left(-\int_{t_{LoS}}^{\infty}\left(1-\exp\left(\mathrm{\mathrm{i\omega\frac{P_{d}A_{BL}v^{-\alpha_{BL}}}{P_{0}\left(A_{BL}r^{-\alpha^{BL}}\right)^{1-\varepsilon}}}}\right)\right)\lambda_{tu}^{L}(v)dv\right)\nonumber \\
\times & \exp\left(-\int_{t_{LoS}}^{\infty}\left(1-\exp\left(\mathrm{i\omega\frac{P_{d}A_{BN}v^{-\alpha_{BN}}}{P_{0}\left(A_{BL}r^{-\alpha^{BL}}\right)^{1-\varepsilon}}}\right)\right)\lambda_{tu}^{NL}(v)dv\right)\nonumber \\
\times & \exp\left(\mathrm{i\omega\frac{\sigma_{c}^{2}}{P_{0}\left(A_{BL}r^{-\alpha^{BL}}\right)^{1-\varepsilon}}}\right),\label{eq:cellular}
\end{align}
and
\begin{align}
\mathcal{F}_{\frac{\textrm{1}}{SINR^{NL}}}(\omega) & =\exp\left(-\int_{r}^{\infty}\left(1-\int_{0}^{t_{NLoS}}\left[\exp\left(\mathrm{i\omega\frac{\mathrm{\left(\frac{z^{\alpha_{BL}}}{A_{BL}}\right)^{\varepsilon}}A_{BL}v^{-\alpha_{BL}}}{\left(A_{BN}r^{-\alpha^{BN}}\right)^{1-\varepsilon}}}\right)\right]f_{\overline{R_{NLCU}}}(z)dz\right)\lambda_{B}^{L}(v)dv\right)\nonumber \\
\times & \exp\left(-\int_{r}^{\infty}\left(1-\int_{0}^{t_{NLoS}}\left[\exp\left(\mathrm{i\omega\frac{\mathrm{\left(\frac{z^{\alpha_{BL}}}{A_{BL}}\right)^{\varepsilon}}A_{BN}v^{-\alpha_{BN}}}{\left(A_{BN}r^{-\alpha^{BN}}\right)^{1-\varepsilon}}}\right)\right]f_{\overline{R_{NLCU}}}(z)dz\right)\lambda_{B}^{NL}(v)dv\right)\nonumber \\
\times & \exp\left(-\int_{t_{NLoS}}^{\infty}\left(1-\exp\left(\mathrm{\mathrm{i\omega\frac{P_{d}A_{BL}v^{-\alpha_{BL}}}{P_{0}\left(A_{BN}r^{-\alpha^{BN}}\right)^{1-\varepsilon}}}}\right)\right)\lambda_{tu}^{L}(v)dv\right)\nonumber \\
\times & \exp\left(-\int_{t_{NLoS}}^{\infty}\left(1-\exp\left(\mathrm{i\omega\frac{P_{d}A_{BN}v^{-\alpha_{BN}}}{P_{0}\left(A_{BN}r^{-\alpha^{BN}}\right)^{1-\varepsilon}}}\right)\right)\lambda_{tu}^{NL}(v)dv\right)\nonumber \\
\times & \exp\left(\mathrm{i\omega\frac{\sigma_{c}^{2}}{P_{0}\left(A_{BN}r^{-\alpha^{BN}}\right)^{1-\varepsilon}}}\right)\label{eq:FUNCTIION SINRNLOS}
\end{align}

\begin{IEEEproof}
See Appendix B.
\end{IEEEproof}
From~\cite{our_work_TWC2016}, $T_{c}^{{\rm {L}}}$ and $T_{c}^{{\rm {NL}}}$
are independent of each other. The coverage probability evaluated
by Eq.(\ref{eq:Theorem_1_p_cov}) is at least a 4-fold integral which
is complicated for numerical computation. However, it gives general
results that can be applied to various multi-path fading or shadowing
model, e.g., Rayleigh fading, Nakagami-m fading, etc, and various
NLoS/LoS transmission models as well.

The third and forth row in Eq.(\ref{eq:cellular}) and Eq.(\ref{eq:FUNCTIION SINRNLOS})
are the aggregate interference from D2D tier. When the mode selection
threshold $\beta$ increases, we can find the intensity of D2D transmitter
also increases. This will reduce the coverage probability performance
of cellular tier, so we make $p_{c}^{{\rm {cov}}}>\varepsilon$ as
a condition to guarantee the performance for cellular mode when choosing
$\beta$.

\subsubsection{Coverage probability of the typical UE in the D2D mode\label{subsec:Coverage-Probability-of}}

From~\cite{7147772}, one can see that in order to derive the coverage
probability of a generic D2D UE, we only need to derive the coverage
probability for a typical D2D receiver UE. Similar to the analysis
in subsection~\ref{subsec:Cellular-mode}, we focus on a typical
D2D UE which is located at the origin $o$ and scheduled to receive
data from another D2D UE. Following Slivnyak's theorem for PPP, the
coverage probability result derived for the typical D2D UE holds also
for any generic D2D UE located at any location. In the following,
we present the coverage probability for a typical D2D UE in Theorem~\ref{thm:We-focus-on}.
\begin{thm}
\label{thm:We-focus-on}We focus on a typical D2D UE which is located
at the origin $o$ and scheduled to receive data from another D2D
UE, the probability of coverage $p_{D2D}^{{\rm {cov}}}\left(\lambda,\gamma\right)$
can be derived as

\noindent
\begin{equation}
p_{D2D}^{{\rm {cov}}}\left(\lambda,\gamma\right)=T_{D2D}^{{\rm {L}}}+T_{D2D}^{{\rm {NL}}},\label{eq:Theorem_1_p_cov-1}
\end{equation}
where $T_{D2D}^{{\rm {L}}}=\int_{0}^{\infty}\left(\mathcal{\int_{\mathrm{-}\infty}^{\infty}\mathrm{\left[\frac{1-e^{-i\omega/\gamma}}{2\pi i\omega}\right]\mathcal{F}_{\frac{\textrm{1}}{SINR_{D2D}^{L}}}(\omega)}}d\omega\right)f_{\overline{R_{LD2D}}}(\overline{R_{d,0}})d\overline{R_{d,0}}$,

\noindent $T_{D2D}^{{\rm {NL}}}=\int_{0}^{\infty}\left(\mathcal{\int_{\mathrm{-}\infty}^{\infty}\mathrm{\left[\frac{1-e^{-i\omega/\gamma}}{2\pi i\omega}\right]\mathcal{F}_{\frac{\textrm{1}}{SINR_{D2D}^{NL}}}(\omega)}}d\omega\right)f_{\overline{R_{NLD2D}}}(\overline{R_{d,0}})d\overline{R_{d,0}}$
,\\
 $f_{\overline{R_{LD2D}}}(r)$ and $f_{\overline{R_{NLD2D}}}(r)$
can be calculated from cumulative distribution function (CDF) of $\overline{R}_{d}^{LOS}$
and $\overline{R}_{d}^{NLOS}$ in appendix C. In addition, $\mathcal{F}_{\frac{\textrm{1}}{SINR_{D2D}^{L}}}(\omega)$
and $\mathcal{F}_{\frac{\textrm{1}}{SINR_{D2D}^{NL}}}(\omega)$ are
respectively computed by
\begin{align}
\mathcal{F}_{\frac{\textrm{1}}{SINR_{D2D}^{L}}}(\omega) & =\exp\left(-\int_{0}^{\infty}\left(1-\int_{0}^{t_{LoS}}\left[\exp\left(\mathrm{i\omega\frac{P_{0}\mathcal{\mathrm{\left(\frac{\overline{R}_{i}^{\alpha_{BL}}}{A_{BL}}\right)^{\varepsilon}}}v^{-\alpha_{dL}}}{P_{d}(\overline{R_{d,0}})^{-\alpha_{dL}}}}\right)\right]f_{\overline{R_{LCU}}}(\overline{R}_{i})d\overline{R}_{i}\right)\lambda_{B}^{L}(v)dv\right)\nonumber \\
\times & \exp\left(-\int_{0}^{\infty}\left(1-\int_{0}^{t_{LoS}}\left[\exp\left(\mathrm{i\omega\frac{P_{0}\mathcal{\mathrm{\left(\frac{\overline{R}_{i}^{\alpha_{BL}}}{A_{BL}}\right)^{\varepsilon}}}A_{DN}v^{-\alpha_{dN}}}{P_{d}A_{DL}(\overline{R_{d,0}})^{-\alpha_{dL}}}}\right)\right]f_{\overline{R_{LCU}}}(\overline{R}_{i})d\overline{R}_{i}\right)\lambda_{B}^{NL}(v)dv\right)\nonumber \\
\times & \exp\left(-\int_{r}^{\infty}\left(1-\exp\left(\mathrm{\mathrm{i\omega\frac{v^{-\alpha_{dL}}}{(\overline{R_{d,0}})^{-\alpha_{dL}}}}}\right)\right)\lambda_{tu}^{L}(v)dv\right)\nonumber \\
\times & \exp\left(-\int_{r}^{\infty}\left(1-\exp\left(\mathrm{i\omega\frac{A_{DN}v^{-\alpha_{dN}}}{A_{DL}(\overline{R_{d,0}})^{-\alpha_{dL}}}}\right)\right)\lambda_{tu}^{NL}(v)dv\right)\nonumber \\
\times & \exp\left(\mathrm{i\omega\frac{\sigma_{d}^{2}}{P_{d}A_{DL}(\overline{R_{d,0}})^{-\alpha_{dL}}}}\right),
\end{align}
and
\begin{align}
\mathcal{F}_{\frac{\textrm{1}}{SINR_{D2D}^{NL}}}(\omega) & =\exp\left(-\int_{0}^{\infty}\left(1-\int_{0}^{t_{NLoS}}\left[\exp\left(\mathrm{i\omega\frac{P_{0}\mathcal{\mathrm{\left(\frac{\overline{R}_{i}^{\alpha_{BL}}}{A_{BL}}\right)^{\varepsilon}}}A_{DL}v^{-\alpha_{dL}}}{P_{d}A_{DN}(\overline{R_{d,0}})^{-\alpha_{dN}}}}\right)\right]f_{\overline{R_{NLCU}}}(\overline{R}_{i})d\overline{R}_{i}\right)\lambda_{B}^{L}(v)dv\right)\nonumber \\
\times & \exp\left(-\int_{0}^{\infty}\left(1-\int_{0}^{t_{NLoS}}\left[\exp\left(\mathrm{i\omega\frac{P_{0}\mathcal{\mathrm{\left(\frac{\overline{R}_{i}^{\alpha_{BL}}}{A_{BL}}\right)^{\varepsilon}}}v^{-\alpha_{dN}}}{P_{d}(\overline{R_{d,0}})^{-\alpha_{dN}}}}\right)\right]f_{\overline{R_{NLCU}}}(\overline{R}_{i})d\overline{R}_{i}\right)\lambda_{B}^{NL}(v)dv\right)\nonumber \\
\times & \exp\left(-\int_{r}^{\infty}\left(1-\exp\left(\mathrm{\mathrm{i\omega\frac{A_{DL}v^{-\alpha_{dL}}}{A_{DN}(\overline{R_{d,0}})^{-\alpha_{dN}}}}}\right)\right)\lambda_{tu}^{L}(v)dv\right)\nonumber \\
\times & \exp\left(-\int_{r}^{\infty}\left(1-\exp\left(\mathrm{i\omega\frac{v^{-\alpha_{dN}}}{(\overline{R_{d,0}})^{-\alpha_{dN}}}}\right)\right)\lambda_{tu}^{NL}(v)dv\right)\nonumber \\
\times & \exp\left(\mathrm{i\omega\frac{\sigma_{d}^{2}}{P_{d}A_{DN}(\overline{R_{d,0}})^{-\alpha_{dN}}}}\right),\label{eq:FUNCTIION SINRNLOS-1}
\end{align}
where $A_{DL}=10^{\frac{1}{10}A_{DL}^{\textrm{dB}}}$ and $A_{DN}=10^{\frac{1}{10}A_{DN}^{\textrm{dB}}}$
denote a constant determined by the transmission frequency for UE-to-UE
links in LoS and NLoS, respectively.
\end{thm}
\begin{IEEEproof}
See Appendix C.
\end{IEEEproof}
The coverage probability of D2D users is evaluated by Eq.(\ref{eq:Theorem_1_p_cov-1}).
Here, we assumed that D2D users are independently distributed regard
to cellular users~\cite{7147772}, so the D2D users follow a Possion
point process. Although the analytical results are complicated, it
provides general results that can be applied to various multi-path
fading or shadowing models in the D2D-enhanced networks.

\section{Simulation and Discussion\label{sec:Simulation-and-Discussion}}

In this section, we use numerical results to validate our results
and analyze the performance of the D2D-enabled UL cellular network.
To this end, we present the simulation parameters, the results for
the coverage probability, the results for the area spectral efficiency
in Section~\ref{subsec:Simulation-Setup},~\ref{subsec:The-Results-on},~\ref{subsec:The-Results-on-1},
respectively.

\subsection{Simulation setup\label{subsec:Simulation-Setup}}

According to the 3GPP LTE specifications~\cite{TR36.872}, we set
the system bandwith to 10MHz, carrier frequency $f_{c}$ to 2GHz,
the BS intensity to $\lambda_{B}=5\,\textrm{BSs/km}^{2}$, which results
in an average inter-site distance of about 500$\,$m. The UE intensity
is chosen as $\lambda=200\,\textrm{UEs/km}^{2}$, which is a typical
value in 5G~\cite{our_work_TWC2016}. The transmit power of each
BS and each D2D transmitter are set to $P_{B}=46\,\textrm{dBm}$ and
$P_{D}=10\,\textrm{dBm}$, respectively. Moreover, the threshold for
selecting cellular mode communication is $\beta=-70\sim-30\textrm{dBm}$.
The standard deviation of lognormal shadowing is $8\,\textrm{dB}$
between UEs to BSs and $7\,\textrm{dB}$ between UEs to UEs. The noise
powers are set to $-95\,\textrm{dBm}$ for a UE receiver and $-114\,\textrm{dBm}$
for a BS receiver, respectively. The simulation parameters are summarized
in Table~\ref{table1}.

\begin{table}

\caption{Simulation Parameters}
\label{table1}
\centering{}%
\begin{tabular}{|c|c|c|c|}
\hline
Parameters & Values & Parameters & Values\tabularnewline
\hline
\hline
$\mathtt{BW}$ & 10MHz & $f_{c}$ & 2GHz\tabularnewline
\hline
$\lambda_{B}$ & {\small{}5 BSs/$km^{2}$} & $\sigma_{c}^{2}$ & {\small{}-95 dBm}\tabularnewline
\hline
$\lambda_{u}$ & {\small{}200 UEs/$km^{2}$} & $\sigma_{d}^{2}$ & {\small{}-114 dBm}\tabularnewline
\hline
$\varepsilon$ & {\small{}0.8} & $P_{0}$ & {\small{}-70 dBm}\tabularnewline
\hline
$\alpha_{BL}$ & {\small{}2.42} & $A_{BL}$ & {\small{}$10^{-3.08}$}\tabularnewline
\hline
$\alpha_{BN}$ & {\small{}4.28} & $A_{BN}$ & {\small{}$10^{-0.27}$}\tabularnewline
\hline
$\alpha_{dL}$ & {\small{}2} & $A_{DL}$ & {\small{}$10^{-3.845}$}\tabularnewline
\hline
$\alpha_{dN}$ & {\small{}4} & $A_{DN}$ & {\small{}$10^{-5.578}$}\tabularnewline
\hline
$P_{b}$ & {\small{}46 dBm} & $P_{d}$ & {\small{}10 dBm}\tabularnewline
\hline
$d_{B}$ & 0.3km & $d_{D}$ & 0.1km\tabularnewline
\hline
\end{tabular}
\end{table}

\subsection{Validation of analytical results of $p^{{\rm {cov}}}\left(\lambda,\gamma\right)$\label{subsec:The-Results-on}}

\begin{figure}
\begin{centering}
\includegraphics[width=12cm]{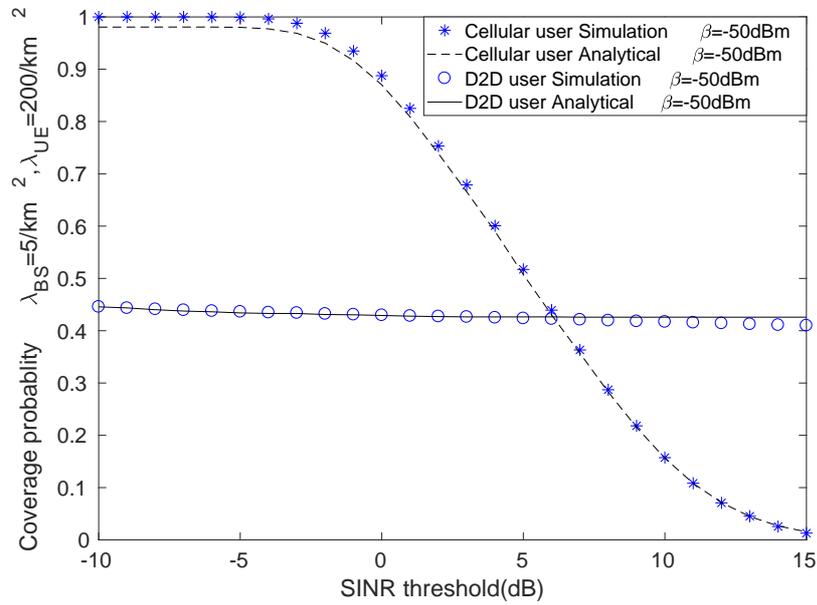}\caption{\label{fig:The-Coverage-Probability}The Coverage Probability $p^{{\rm {cov}}}\left(\lambda,\gamma\right)$
vs. SINR threshold ($\lambda_{UE}=200\,\textrm{UEs/km}^{2}$, $\lambda_{BS}=5\,\textrm{UEs/km}^{2}$
and $\rho=10\%$). The mode select threshold $\beta$ is $-50\text{dBm}$. }
\par\end{centering}
\end{figure}

In Fig.\ \ref{fig:The-Coverage-Probability}, we plot the results
of the coverage probability of cellular tier and D2D tier, we can
draw the following observations:
\begin{itemize}
\item The analytical results of the coverage probability from Eq.(\ref{eq:Theorem_1_p_cov})
and Eq.(\ref{eq:Theorem_1_p_cov-1}) match well with the simulation
results, which validates our analysis and shows that the adopted model
accurately captures the features of D2D communications.
\item The coverage probability decreases with the increase of SINR threshold,
because a higher SINR requirement makes it more difficult to satisfy
the coverage criterion in Eq.(\ref{eq:definesinr-1}).
\item For D2D tier, the coverage probability reduces very slowly because
the signals in most of the successful links are LoS while the interference
is most likely NLoS, hence the SINR is relatively large, e.g., well
above 15 dB.
\end{itemize}
\begin{figure}[h]
\begin{centering}
\includegraphics[width=12cm]{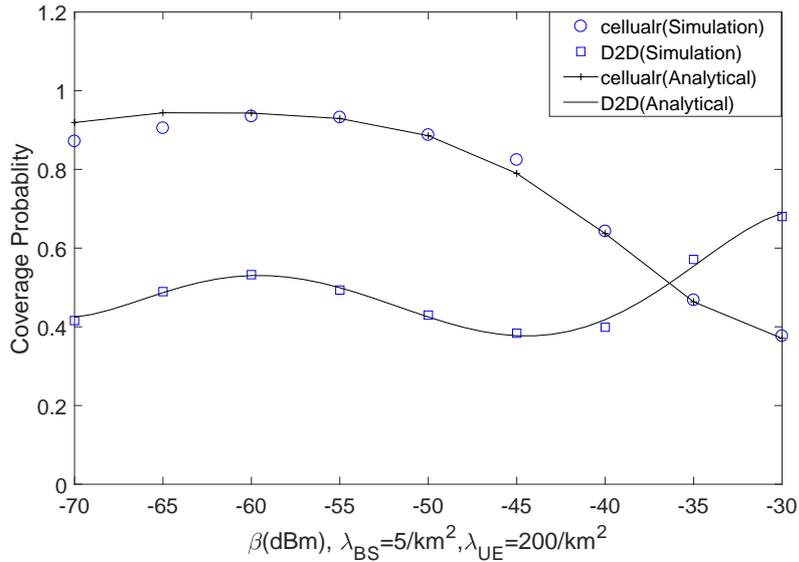}
\par\end{centering}
\caption{\label{fig:The-Coverage-Probability-1}The Coverage Probability $p^{{\rm {cov}}}\left(\lambda,\gamma\right)$
vs. $\beta$ for 3GPP Case~1 ($\gamma_{0}=0\,\textrm{dB}$, $\lambda_{UE}=200\,\textrm{UEs/km}^{2}$,
$\lambda_{BS}=5\,\textrm{UEs/km}^{2}$ and $\rho=10\%$).}

\end{figure}

To fully study the SINR coverage probability with respect to the values
of $\beta$ , the results of coverage probability with various $\beta$
and $\gamma_{0}$=0 dB are plotted in Fig \ref{fig:The-Coverage-Probability-1}.
From this figure, we can draw the following observations:
\begin{itemize}
\item The coverage probability of cellular users increases as $\beta$ grows
from -70 dBm to -57 dBm, which is because a larger $\beta$ reduces
the distance between the typical CU to the typical BS so that the
signal link's LoS probability increases. Then, the coverage probability
performance decreases because the interference from D2D tier is growing.
When we set $\varepsilon=0.9$, we should choose $\beta$ no larger
than -45 dBm to guarantee the cellular performance.
\item In the D2D mode, the coverage probability also increases as $\beta$
increases from -70 dBm to -60 dBm, this is because the distance between
the typical D2D pair UEs decreases while the transmit power is constant.
From $\beta=-60$ dBm to $\beta=-45$ dBm, the coverage probability
decreases because the interference from the D2D tier increases. Then,
the coverage probability increases when $\beta$ is larger than -45
dBm because the signal power experience the NLoS to LoS transition
while the aggregate interference remains to be mostly NLoS interference.
\end{itemize}

\subsection{Discussion on the analytical results of ASE\label{subsec:The-Results-on-1}}

\begin{figure}[h]
\begin{centering}
\includegraphics[width=12cm]{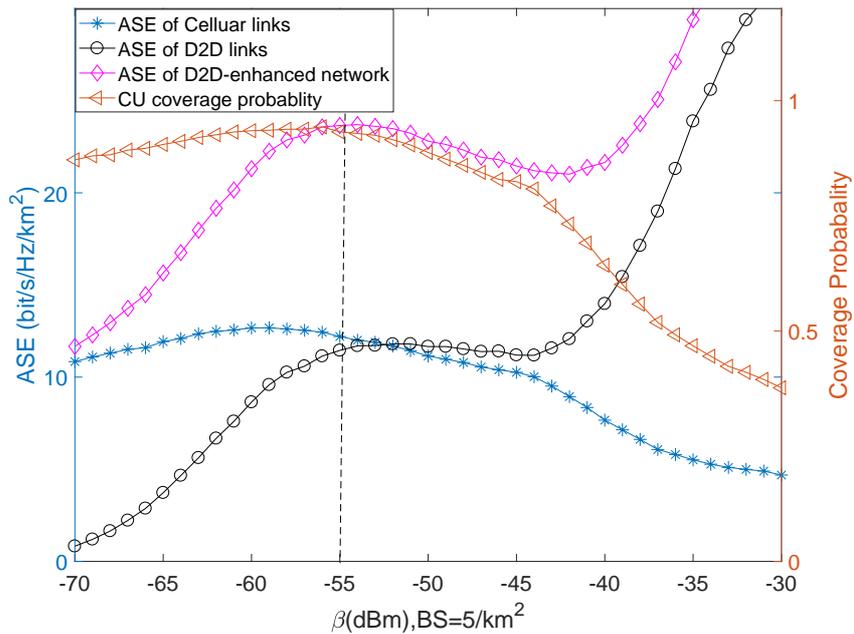}\caption{\label{fig:The-ASE-}The ASE $A^{\textrm{ASE}}\left(\lambda,\gamma_{0}\right)$
vs. $\beta$ for 3GPP Case~1 ($\gamma_{0}=0\,\textrm{dB}$, $\lambda_{UE}=200\,\textrm{UEs/km}^{2}$,
$\lambda_{BS}=5\,\textrm{UEs/km}^{2}$ and $\rho=10\%$).}
\par\end{centering}
\end{figure}

The analytical results of ASE with $\gamma_{0}$=0 db vs various $\beta$
values are shown in Eq.(\ref{eq:ase}). Fig.\ref{fig:The-ASE-} illustrates
the ASEs of Cellular links, D2D links and of the whole network with
respect to different mode selection thresholds $\beta$ . From this
figure we can draw the following observations:
\begin{itemize}
\item The total ASE increases when $\beta\in[-55dBm,-42dBm]$, as the D2D
links increases, because they do not generate a lot of interference
to the cellular tier.
\item An optimal $\beta$ around$-55$ dBm can achieve the maximum ASE while
the coverage probability of the cellular tier is above 0.9.
\item When $\beta\in[-55dBm,-42dBm]$, the total ASE decreases because the
D2D links generate more interference which makes the coverage probability
of cellular UEs suffer. The ASE and the coverage probability of cellular
links also decrease because the aggregate interference is now mostly
LoS interference.
\item When $\beta\in[-42dBm,-30dBm]$, the additional D2D links make significant
contribution to the ASE performance so that the total ASE grows again.
Then, the total ASE approaches that of the D2D ASE because the percentage
of D2D UE is approaching 100\%, which has been analyzed in Eq.(\ref{eq:q}).
Although the total ASE grows very quickly when $\beta\in[-42dBm,-30dBm]$,
the interference from D2D links to the cellular tier remains to be
large so that the performance of the cellular tier is poor. Hence,
we do not recommend the network operate in this range of $\beta$.
\end{itemize}
From Fig.\ref{fig1} we can find D2D links will increase as $\beta$
increase for all different densities of BS. At first, D2D links will
enhance the ASE performance but they do not generate a lot of interference
to the cellular tier. Then the increase of D2D transmitter will generate
more interference which makes the coverage probability of cellular
UEs suffer. The optimal $\beta$ can be find in this stage for different
densities of BS. At last the total ASE approaches to that of the D2D
ASE because the percentage of D2D UE is approaching 100\%. Above all,
there exists an optimal $\beta$ that can achieve the maximum ASE
of the D2D-enabled cellular while the coverage probability in cellular
tier is guaranteed. The mode selection threshold can control the interference
from both cellular tier and D2D tier. In addition, the D2D tier can
nearly double the ASE for the network when appropriately choosing
the threshold for mode selection.

\section{Conclusion\label{sec:Conclusion}}

In this paper, we proposed an interference management method in a
D2D enhanced uplink cellular network, where the location of the mobile
UEs and the BSs are modeled as PPPs. In particular, each UE selects
its operation mode based on its downlink received power and an interference
threshold $\beta$. Practical pathloss and slow shadow fading are
consider in modeling the power attenuation. This mode selection method
mitigates large interference from D2D transmitter to cellular network.
Using a stochastic geometric approach, we analytically evaluated the
coverage probability and the ASE for various values of the mode selection
threshold $\beta$. Our results showed that the D2D links can provide
high ASE when the threshold parameter is appropriately chosen. More
importantly, we concluded that there exists an optimal $\beta$ to
achieve the maximum ASE while guaranteeing the coverage probability
performance of the cellular network.

As our future work, we will consider other factors of realistic networks
in the theoretical analysis for SCNs, such as practical directional
antennas~\cite{7126919} and non-HPPP deployments of BSs~\cite{7959926}.

\section*{Appendix\ A:Proof of Lemma\ \ref{lem:When-operating-under}\label{sec:AppendixA:Proof-of-Lemma}}
\begin{IEEEproof}
\noindent The probability that the RSS is larger than the threshold
is given by
\begin{equation}
P=\Pr\left[\underset{b}{\max}\left\{ P_{b}^{\textrm{rx}}\right\} >\beta\right],\label{eq:receivepower}
\end{equation}
where we use the standard power loss propagation model with a path
loss exponent $\alpha_{BL}$ (for LoS UE-BS links) and $\alpha_{BN}$
(for NLoS UE-BS links).The probability that a generic mobile UE operates
in the cellular mode is
\begin{eqnarray}
q & = & \mathsf{\mathit{\mathrm{1}-\Pr\left[\underset{b}{\max}\left\{ P_{b}^{\textrm{rx}}\right\} \leq\beta\right]}}\nonumber \\
 & = & 1-\Pr\left[\max\left\{ P_{LOS}^{rx}\right\} \leq\beta\cap\max\left\{ P_{NLOS}^{rx}\right\} \leq\beta\right]\nonumber \\
 & = & 1-\Pr\left[\min\overline{R}_{i}^{BL}\geq\left(\frac{P_{b}\textrm{A}_{BL}}{\beta}\right)^{1/\alpha_{BL}}\cap\min\overline{R}_{i}^{BN}\geq\left(\frac{P_{b}\textrm{A}_{BN}}{\beta}\right)^{1/\alpha_{BN}}\right]\nonumber \\
 & = & 1-\Pr\left[\textrm{no nodes within \ensuremath{\left(\frac{P_{b}\textrm{A}_{BL}}{\beta}\right)^{1/\alpha_{BL}}}}\cap\textrm{no nodes within \ensuremath{\left(\frac{P_{b}\textrm{A}_{BN}}{\beta}\right)^{1/\alpha_{BN}}}}\right]\nonumber \\
 & = & 1-\exp\left[-\wedge^{\textrm{NL}}\left(\left[0,\left(\frac{P_{b}\textrm{A}_{BL}}{\beta}\right)^{1/\alpha_{BL}}\right]\right)\right]\cdot\exp\left[-\wedge^{\textrm{L}}\left[0,\left(\frac{P_{b}\textrm{A}_{BN}}{\beta}\right)^{1/\alpha_{BN}}\right]\right]\nonumber \\
 & = & 1-\exp\left[-\mathbb{E}_{\mathcal{H}}\left[2\pi\lambda\int_{0}^{\left(\frac{P_{b}\textrm{A}_{BL}\mathcal{H}}{\beta}\right)^{1/\alpha_{BL}}}p^{L}(r)rdr\right]\right]\nonumber \\
 &  & \cdot\exp\left[-\mathbb{E}_{\mathcal{H}}\left[2\pi\lambda\int_{0}^{\left(\frac{P_{b}\textrm{A}_{BN}\mathcal{H}}{\beta}\right)^{1/\alpha_{BN}}}p^{NL}(r)rdr\right]\right],
\end{eqnarray}
which concludes our proof.
\end{IEEEproof}

\section*{Appendix B:Proof of Theorem 2}
\begin{IEEEproof}
By invoking the law of total probability, the coverage probability
of cellular links can be divided into two parts, i.e., $T_{c}^{{\rm {L}}}+T_{c}^{{\rm {NL}}}$,
which denotes the conditional coverage probability given that the
typical BS is associated with a BS in LoS and NLoS, respectively.
First, we derive the coverage probability for LoS link cellular tier.
Conditioned on the strongest BS being at a distance $R_{B,0}$ from
the typical CU, the equivalence distance$\overline{R_{LOSCU}}=\mathrm{\mathcal{H}_{B}^{-1/\alpha_{BL}}R_{B,0}}$
$\left(\overline{R_{LOSCU}}\leq\left(\frac{\beta}{\textrm{\ensuremath{P_{B}A^{L}}}}\right){}^{-1/\alpha_{BL}}\right)$,
probability of coverage is given by
\begin{align}
T^{{\rm {L}}} & =\Pr\left[\frac{1}{SINR^{L}}<\frac{1}{\gamma}\left|\textrm{LOS}\right.\right]\nonumber \\
= & \int_{0}^{t_{LoS}}\left(\int_{0}^{\frac{1}{\gamma}}f_{\frac{\textrm{1}}{SINR^{L}}}\left(x\right)dx\right)f_{\overline{R_{LCU}}}(r)dr\nonumber \\
= & \int_{0}^{t_{LoS}}\left(\int_{0}^{\frac{1}{\gamma}}\frac{1}{2\pi}\mathcal{\int_{\mathrm{-}\infty}^{\infty}F}_{\frac{\textrm{1}}{SINR^{L}}}(\omega)\cdot e^{-iwx}\cdot d\omega dx\right)f_{\overline{R_{LCU}}}(r)dr\nonumber \\
= & \int_{0}^{t_{LoS}}\left(\mathcal{\int_{\mathrm{-}\infty}^{\infty}\mathrm{\left[\frac{1-e^{-i\omega/\gamma}}{2\pi i\omega}\right]\mathcal{F}_{\frac{\textrm{1}}{SINR^{L}}}(\omega)}}d\omega\right)f_{\overline{R_{LCU}}}(r)dr\text{.}
\end{align}
where $i=\sqrt{-1}$ is the imaginary unit. The inner integral is
the conditional PDF of $\frac{1}{SINR}$; The intensity of cellular
UEs and D2D UEs can be calculated as
\begin{align}
\lambda_{B}^{L}(t) & =2\pi\lambda_{b}\frac{d}{dt}\left(\int_{0}^{\infty}\left[\int_{0}^{t\left(\mathcal{H}\right)^{1/\alpha^{L}}}\Pr^{L}(r)rdr\right]f\left(H\right)dH\right),
\end{align}

\begin{align}
\lambda_{B}^{NL}(t) & =\frac{d}{dt}\left(\mathbb{E}_{\mathcal{H}}\left[2\pi\lambda_{b}\int_{0}^{t\left(\mathcal{H}\right)^{1/\alpha^{NL}}}\Pr^{NL}(r)rdr\right]\right),
\end{align}

\begin{align}
\lambda_{tu}^{L}(t) & =\frac{d}{dt}\left(\mathbb{E}_{\mathcal{H}}\left[\pi\left(1-q\right)\lambda_{u}\int_{0}^{t\left(\mathcal{H}\right)^{1/\alpha^{L}}}\Pr^{L}(r)rdr\right]\right),
\end{align}

\begin{align}
\lambda_{tu}^{NL}(t) & =\frac{d}{dt}\left(\mathbb{E}_{\mathcal{H}}\left[\pi\left(1-q\right)\lambda_{u}\int_{0}^{t\left(\mathcal{H}\right)^{1/\alpha^{NL}}}\Pr^{NL}(r)rdr\right]\right),
\end{align}
$\mathcal{F}_{SINR^{-1}}(\omega)$ denotes the conditional characteristic
function of $\frac{1}{SINR}$, which can be written by

$\mathcal{F}_{\frac{\textrm{1}}{SINR^{L}}}(\omega)$
\begin{eqnarray}
 & = & \int_{R^{2}}f_{\frac{\textrm{1}}{SINR^{L}}}\left(x\right)e^{i\omega x}dx\nonumber \\
 & = & E_{\Phi}\left[\exp\left(\mathrm{i\omega\frac{1}{SINR^{L}}}\right)\left|R_{typicalcu}=\overline{r}\right.\right]\nonumber \\
 & = & \mathbb{E_{\Phi}\left[\exp\left(\mathrm{i\omega\frac{I_{c}+I_{d}+\sigma^{2}}{S^{L}}}\right)\mathrm{\left|R_{typicalcu}=\overline{r}\right.}\right]}\nonumber \\
 & = & \mathbb{E}_{\Phi}\left[\left.\exp\left(\mathrm{i\omega\frac{I_{c}}{S^{L}}}\right)\exp\left(\mathrm{i\omega\frac{I_{d}}{S^{L}}}\right)\exp\left(\mathrm{i\omega\frac{\sigma^{2}}{S^{L}}}\right)\right|R_{typicalcu}=\overline{r}\right].
\end{eqnarray}
By applying stochastic geometry and the probability generating functional(PGFL)
of the PPP. $\mathcal{F}_{\frac{\textrm{1}}{SINR^{L}}}(\omega)$ can
be written as three parts, namely $\mathcal{L_{\mathrm{I_{c}}}\mathrm{(\omega)}}$,$\mathcal{L}_{\mathrm{I_{d}}}\mathrm{(\omega)}$
and $\mathcal{L_{\mathrm{n}}\mathrm{(\omega)}}$,

\begin{align}
\mathcal{L}_{\mathrm{I_{c}}}\mathrm{(\omega)=} & \exp\left(\mathrm{i\omega\frac{I_{CL}+I_{CN}}{S^{L}}}\right)\nonumber \\
= & \exp\left(-\int_{r}^{\infty}\left(1-\int_{0}^{t_{LoS}}\left[\exp\left(\mathrm{i\omega\frac{\mathrm{\left(z^{\alpha_{BL}}\right)^{\varepsilon}}v^{-\alpha_{BL}}}{A_{BL}^{2\epsilon}\left(r^{-\alpha^{BL}}\right)^{1-\varepsilon}}}\right)\right]f_{\overline{R_{LCU}}}(z)dz\right)\lambda_{B}^{L}(v)dv\right.\nonumber \\
 & \left.-\int_{r}^{\infty}\left(1-\int_{0}^{t_{LoS}}\left[\exp\left(\mathrm{i\omega\frac{\mathcal{\mathrm{\left(z^{\alpha_{BL}}\right)^{\varepsilon}}}v^{-\alpha_{BN}}}{A_{BL}^{2\epsilon}\left(r^{-\alpha^{BL}}\right)^{1-\varepsilon}}}\right)\right]f_{\overline{R_{LCU}}}(z)dz\right)\lambda_{B}^{NL}(v)dv\right),
\end{align}
and
\begin{align}
\mathcal{L}_{\mathrm{I_{d}}}\mathrm{(\omega)=} & \exp\left(\mathrm{i\omega\frac{I_{DL}+I_{DN}}{S^{L}}}\right)\nonumber \\
= & \exp\left(-\int_{t_{LoS}}^{\infty}\left(1-\exp\left(\mathrm{\mathrm{i\omega\frac{P_{d}A_{BL}v^{-\alpha_{BL}}}{P_{0}\left(A_{BL}r^{-\alpha^{BL}}\right)^{1-\varepsilon}}}}\right)\right)\lambda_{tu}^{L}(v)dv\right.\nonumber \\
 & \left.-\int_{t_{LoS}}^{\infty}\left(1-\exp\left(\mathrm{i\omega\frac{P_{d}A_{BN}v^{-\alpha_{BN}}}{P_{0}\left(A_{BL}r^{-\alpha^{BL}}\right)^{1-\varepsilon}}}\right)\right)\lambda_{tu}^{NL}(v)dv\right),
\end{align}
and$\mathcal{L_{\mathrm{n}}\mathrm{(\omega)}}=\exp\left(\mathrm{iw\frac{\sigma^{2}}{P_{0}\left(A_{BL}r^{-\alpha^{BL}}\right)^{1-\varepsilon}}}\right)$
which is the cellular interference , D2D interference and noise part
in characteristic function.

Finally, note that the value of $p_{c}^{{\rm {cov}}}\left(\lambda,\gamma\right)$
in Eq. (\ref{eq:Theorem_1_p_cov}) should be calculated by taking
the expectation with $f_{\overline{R_{LCU}}}(r)$ and $f_{\overline{R_{NLCU}}}(r)$,
which is given as follow
\begin{align}
f_{\overline{R_{LCU}}}(r) & =\left(\frac{d}{dr}\left\{ 1-\exp\left[-\varLambda^{L}\left(\left[0,r\right]\right)\right]\cdot\exp\left[-\varLambda^{NL}\left(\left[0,\overline{r_{1}}\right]\right)\right]\right\} \left|CU\right.\right)\nonumber \\
 & =\frac{\exp\left[-\varLambda^{L}\left(\left[0,r\right]\right)\right]\cdot\exp\left[-\varLambda^{NL}\left(\left[0,\overline{r_{1}}\right]\right)\right]{\rm {Pr}}^{{\rm {L}}}\left(r\right)\lambda_{B}^{L}(r)}{q},
\end{align}
where the typical UE should guarantee that there is no NLoS BS in
$\overline{r_{1}}$ when the signal is LoS. Given that the typical
BS is connected to a NLoS UE, the conditional coverage probability
$T^{{\rm {N}}}$ can be derived in a similar way as the above. In
this way, the coverage probability is obtained by $T_{c}^{{\rm {L}}}+T_{c}^{{\rm {NL}}}$.
Which concludes our proof.
\end{IEEEproof}

\section*{Appendix C:Proof of Theorem 3}
\begin{IEEEproof}
\label{lem:The-typical-D2D cdf of d2d distance}The typical D2D receiver
selects the equivalent nearest UE as a potential transmitter. If the
potential D2D receiver is operating in a cellular mode, D2D RU must
search for another transmitter. We approximately consider that the
second neighbor can be found as the transmitter under this situation
both for LoS/NLoS links. The approximate cumulative distribution function(CDF)
of $\overline{R}_{d}^{LOS}$ can be written as
\begin{align}
\Pr\left[\overline{R}_{d}^{LOS}<R\right]\nonumber \\
\approx & \int_{R+t_{LoS}}^{\infty}\left(\int_{0}^{R}f_{R_{d}^{LOS}}(\overline{R}_{d})d\overline{R}_{d}\right)f_{r_{1}^{LOS}}(r_{1})dr_{1}\nonumber \\
+ & \int_{t_{LoS}}^{R+t_{LoS}}\left(\int_{0}^{r_{1}-t_{LoS}}f_{R_{d}}(\overline{R}_{d})d\overline{R}_{d}\right.\nonumber \\
+ & \int_{r_{1}-t_{LoS}}^{R}(1-P_{c}^{L})\cdot f_{R_{d}^{LOS}}(\overline{R}_{d})d\overline{R}_{d}\nonumber \\
+ & \left.\int_{r_{1}-t_{LoS}}^{R}P_{c}^{L}\cdot f_{R_{d_{2}}^{LOS}}\left(\overline{R}_{d}\right)d\overline{R}_{d}\right)f_{r_{1}^{LOS}}(r_{1})dr_{1}\nonumber \\
+ & \int_{R+t_{NLoS}}^{\infty}\left(\int_{0}^{R}f_{R_{d}^{LOS}}(\overline{R}_{d})d\overline{R}_{d}\right)f_{r_{1}^{NLOS}}(r_{1})dr_{1}\nonumber \\
+ & \int_{t_{NLoS}}^{R+t_{NLoS}}\left(\int_{0}^{r_{1}-t}f_{R_{d}^{LOS}}(\overline{R}_{d})d\overline{R}_{d}\right.\nonumber \\
+ & \int_{r_{1}-t_{NLoS}}^{R}(1-P_{c}^{L})\cdot f_{R_{d}^{LOS}}(\overline{R}_{d})d\overline{R}_{d}\nonumber \\
+ & \left.\int_{r_{1}-t_{NLoS}}^{R}P_{c}^{L}\cdot f_{R_{d_{2}}^{LOS}}\left(\overline{R}_{d}\right)d\overline{R}_{d}\right)f_{r_{1}^{NLOS}}(r_{1})dr_{1},\label{eq:pdf of los d2dlink}
\end{align}
where $r_{1}$ is the equivalent distance from TU to the strongest
LoS/NLoS BS, $t_{LoS}=\left(\frac{\beta}{\textrm{\ensuremath{P_{B}A^{L}}}}\right){}^{-1/\alpha_{BL}}$,$t_{NLoS}=\left(\frac{\beta}{\textrm{\ensuremath{P_{B}A^{NL}}}}\right){}^{-1/\alpha_{BN}}$,
$P_{c}^{L/NL}$is the probability of a D2D receiver be a CU.
\begin{equation}
f_{r_{1}^{LOS}}(r)=\frac{\exp\left[-\varLambda^{L}\left(\left[0,r\right]\right)\right]\cdot\exp\left[-\varLambda^{NL}\left(\left[0,\overline{r_{1}}\right]\right)\right]{\rm {Pr}}_{{\rm {B}}}^{{\rm {L}}}\left(r\right)\lambda_{B}^{L}(r)}{1-q}\label{eq:distane D2D to LOS bs}
\end{equation}
and
\begin{equation}
f_{r_{1}^{NLOS}}(r)=\frac{\exp\left[-\varLambda^{NL}\left(\left[0,r\right]\right)\right]\cdot\exp\left[-\varLambda^{L}\left(\left[0,\overline{r_{1}}\right]\right)\right]{\rm {Pr}}_{{\rm {B}}}^{{\rm {NL}}}\left(r\right)\lambda_{B}^{NL}(r)}{1-q}\label{eq:distane D2D to NLOS bs}
\end{equation}
According to~\cite{our_work_TWC2016}, if there is no difference
between CUs and D2D UEs, the pdf of the distance for a tier of PPP
LoS UEs is
\begin{equation}
f_{R_{d}^{LOS}}(r)=\exp\left(\hspace{-0.1cm}-\hspace{-0.1cm}\int_{0}^{\overline{r_{1}}}{\rm {Pr}}_{{\rm {D}}}^{{\rm {NL}}}\left(u\right)\lambda_{u}^{NL}(u)du\right)\exp\left(\hspace{-0.1cm}-\hspace{-0.1cm}\int_{0}^{r}{\rm {Pr}}_{{\rm {D}}}^{{\rm {L}}}\left(u\right)\lambda_{u}^{L}(u)du\right){\rm {Pr}}_{{\rm {D}}}^{{\rm {L}}}\left(r\right)\lambda_{u}^{L}(r)
\end{equation}
and if there is no difference between CUs and D2D UEs, the pdf of
the distance for a tier of PPP NLoS UEs is
\begin{equation}
f_{R_{d}^{NLOS}}(r)=\exp\left(\hspace{-0.1cm}-\hspace{-0.1cm}\int_{0}^{\overline{r_{2}}}{\rm {Pr}}_{{\rm {D}}}^{{\rm {L}}}\left(u\right)\lambda_{u}^{L}(u)du\right)\exp\left(\hspace{-0.1cm}-\hspace{-0.1cm}\int_{0}^{r}{\rm {Pr}}_{{\rm {D}}}^{{\rm {NL}}}\left(u\right)\lambda_{u}^{NL}(u)du\right){\rm {Pr}}_{{\rm {D}}}^{{\rm {NL}}}\left(r\right)\lambda_{u}^{NL}(r),
\end{equation}
where
\begin{equation}
\lambda_{u}^{L}(r)=\frac{d}{dt}\left(\mathbb{E}_{\mathcal{H}}\left[2\pi\left(1-q\right)\lambda_{u}\int_{0}^{t\left(\mathcal{H}\right)^{1/\alpha^{L}}}{\rm {Pr}}_{{\rm {D}}}^{{\rm {L}}}(r)rdr\right]\right),
\end{equation}
and
\begin{equation}
\lambda_{u}^{NL}(r)=\frac{d}{dt}\left(\mathbb{E}_{\mathcal{H}}\left[2\pi\left(1-q\right)\lambda_{u}\int_{0}^{t\left(\mathcal{H}\right)^{1/\alpha^{NL}}}{\rm {Pr}}_{{\rm {D}}}^{{\rm {NL}}}(r)rdr\right]\right),
\end{equation}
According to~\cite{1512427} , the second neighbor point is distributed
as
\begin{equation}
f_{R_{d_{2}}^{LOS}}(r)=2\pi^{2}r^{3}\lambda_{u}^{L}(t)^{2}\cdotp\exp\left[-\mathbb{E}_{\mathcal{H}}\left[2\pi\lambda_{u}\int_{0}^{r\left(\mathcal{H}\right)^{1/\alpha^{L}}}{\rm {Pr}}_{{\rm {D}}}^{{\rm {L}}}rdr\right]\right],\label{eq:distance distribution-los}
\end{equation}
and
\begin{equation}
f_{R_{d_{2}}^{NLOS}}(r)=2\pi^{2}r^{3}\lambda_{u}^{NL}(t)^{2}\cdotp\exp\left[-\mathbb{E}_{\mathcal{H}}\left[2\pi\lambda_{u}\int_{0}^{r\left(\mathcal{H}\right)^{1/\alpha^{NL}}}{\rm {Pr}}_{{\rm {D}}}^{{\rm {NL}}}rdr\right]\right],\label{eq:distance distribution-NLOS}
\end{equation}
similarity, the cdf of the distance of NLoS D2D signal can be written
as
\begin{align}
\Pr\left[\overline{R}_{d}^{NLOS}<R\right]\nonumber \\
\approx & \int_{R+t_{LoS}}^{\infty}\left(\int_{0}^{R}f_{R_{d}^{NLOS}}(\overline{R}_{d})d\overline{R}_{d}\right)f_{r_{1}^{LOS}}(r_{1})dr_{1}\nonumber \\
+ & \int_{t_{LoS}}^{R+t_{LoS}}\left(\int_{0}^{r_{1}-t_{LoS}}f_{R_{d}^{NLOS}}(\overline{R}_{d})d\overline{R}_{d}\right.\nonumber \\
+ & \int_{r_{1}-t_{LoS}}^{R}(1-P_{c}^{NL})\cdot f_{R_{d}^{NLOS}}(\overline{R}_{d})d\overline{R}_{d}\nonumber \\
+ & \left.\int_{r_{1}-t_{LoS}}^{R}P_{c}^{NL}\cdot f_{R_{d_{2}}^{NLOS}}\left(\overline{R}_{d}\right)d\overline{R}_{d}\right)f_{r_{1}^{LOS}}(r_{1})dr_{1}\nonumber \\
+ & \int_{R+t_{NLoS}}^{\infty}\left(\int_{0}^{R}f_{R_{d}^{NLOS}}(\overline{R}_{d})d\overline{R}_{d}\right)f_{r_{1}^{NLOS}}(r_{1})dr_{1}\nonumber \\
+ & \int_{t_{NLoS}}^{R+t_{NLoS}}\left(\int_{0}^{r_{1}-t}f_{R_{d}^{NLOS}}(\overline{R}_{d})d\overline{R}_{d}\right.\nonumber \\
+ & \int_{r_{1}-t_{NLoS}}^{R}(1-P_{c}^{NL})\cdot f_{R_{d}^{NLOS}}(\overline{R}_{d})d\overline{R}_{d}\nonumber \\
+ & \left.\int_{r_{1}-t_{NLoS}}^{R}P_{c}^{NL}\cdot f_{R_{d_{2}}^{NLOS}}\left(\overline{R}_{d}\right)d\overline{R}_{d}\right)f_{r_{1}^{NLOS}}(r_{1})dr_{1},\label{eq:pdf of los d2dlink-1}
\end{align}
the pdf of $\overline{R_{d}}^{L(NL)}$ can be written as
\begin{equation}
f_{\overline{R_{d}}^{L(NL)}}(r)=\frac{\partial\Pr\left[R_{d}^{L(NL)}>r\right]}{\partial\overline{R_{d}}},
\end{equation}
where $P_{c}$ is the probability of the potential D2D receiver operating
in the cellular mode, and it can be calculated as
\begin{equation}
P_{c}^{LOS/NLOS}=\arccos\left(\frac{\overline{R}_{d}+r_{1}^{2}-t_{LOS/NLOS}^{2}}{2\overline{R}_{d}r_{1}}\right)/\pi,\label{eq:circle}
\end{equation}
which concludes our proof.
\end{IEEEproof}
\bibliographystyle{unsrt-fr}
\bibliography{reference}

\end{document}